\documentclass{article}

\usepackage{microtype}
\usepackage{graphicx}
\usepackage{subfigure}
\usepackage{booktabs} 

\usepackage{hyperref}

\usepackage[accepted]{ml4astro2025}

\usepackage{amsmath}
\usepackage{amssymb}
\usepackage{mathtools}
\usepackage{amsthm}

\usepackage{soul}

\usepackage[capitalize,noabbrev]{cleveref}

\theoremstyle{plain}

\theoremstyle{definition}

\theoremstyle{remark}

\usepackage[textsize=tiny]{todonotes}

\newcommand{\icmltitle}{\mlforastrotitle}
\newcommand{\icmltitlerunning}{\mlforastrotitlerunning} 
\newcommand{\icmlauthor}{\mlforastroauthor}
\newcommand{\icmlaffiliation}{\mlforastroaffiliation} 
\newcommand{\icmlcorrespondingauthor}{\mlforastrocorrespondingauthor}
\newcommand{\icmlkeywords}{\mlforastrokeywords}

\newcommand{\icmlsetsymbol}{\mlforastrosetsymbol}
\newenvironment{icmlauthorlist}
  {\begin{mlforastroauthorlist}}
  {\end{mlforastroauthorlist}}

\icmltitlerunning{Predicting the Subhalo Mass Functions in Simulations from Galaxy Images}

\begin{document}

\twocolumn[
\icmltitle{Predicting the Subhalo Mass Functions in Simulations from Galaxy Images}

\icmlsetsymbol{equal}{*}

\begin{icmlauthorlist}
\icmlauthor{Andreas Filipp}{UdeM,MILA,CIELA}
\icmlauthor{Tri Nguyen}{CIERA,SKAI}
\icmlauthor{Laurence Perreault-Levasseur}{UdeM,MILA,CIELA,Flatiron,Perimeter,TSI}
\icmlauthor{Jonah Rose}{CCA}
\icmlauthor{Chris Lovell}{Kavli,InstituteOfAstro}
\icmlauthor{Nicolas Payot}{UdeM,MILA,CIELA}
\icmlauthor{Francisco Villaescusa-Navarro}{Princeton,Simons}
\icmlauthor{Yashar Hezaveh}{UdeM,MILA,CIELA,Flatiron,TSI}
\end{icmlauthorlist}

\icmlaffiliation{UdeM}{Department of Physics, University of Montreal, Montreal, Canada}
\icmlaffiliation{MILA}{MILA Quebec AI Institute, Montreal, Canada}
\icmlaffiliation{CIELA}{CIELA Institute, Montreal Institute for Astrophysics and Machine Learning, Montreal, Canada}
\icmlaffiliation{Flatiron}{Center for Computational Astrophysics, Flatiron Institute, New York, USA}
\icmlaffiliation{Perimeter}{Perimeter Institute for Theoretical Physics, Waterloo, Canada}
\icmlaffiliation{TSI}{Trottier Space Institute, McGill University, Montreal, Canada}
\icmlaffiliation{CIERA}{Center for Interdisciplinary Exploration and Research in Astrophysics, Northwestern University, Evanston, USA}
\icmlaffiliation{SKAI}{The NSF-Simons AI Institute for the Sky, Chicago, USA}
\icmlaffiliation{CCA}{Center for Computational Astrophysics, New York, USA}
\icmlaffiliation{Princeton}{Princeton University, Princeton, USA}
\icmlaffiliation{Simons}{Simons Foundation, New York, USA}
\icmlaffiliation{Kavli}{Kavli Institute for Cosmology, Madingley Road, Cambridge, UK}
\icmlaffiliation{InstituteOfAstro}{Institute of Astronomy, Madingley Road, Cambridge, UK}

\icmlcorrespondingauthor{Andreas Filipp}{andreas.filipp@umontreal.ca}

\icmlkeywords{Machine Learning, ICML, Astrophysics, Dark Matter}

\vskip 0.3in
]

\printAffiliationsAndNotice{}  

\begin{abstract}
Strong gravitational lensing provides a powerful tool to directly infer the dark matter (DM) subhalo mass function (SHMF) in lens galaxies. However, comparing observationally inferred SHMFs to theoretical predictions remains challenging, as the predicted SHMF can vary significantly between galaxies — even within the same cosmological model — due to differences in the properties and environment of individual galaxies.
We present a machine learning framework to infer the galaxy-specific predicted SHMF from galaxy images, conditioned on the assumed inverse warm DM particle mass $M^{-1}_{\rm DM}$. To train the model, we use 1024 high-resolution hydrodynamical zoom-in simulations from the DREAMS suite. Mock observations are generated using \texttt{Synthesizer}, excluding gas particle contributions, and SHMFs are computed with the \texttt{Rockstar} halo finder.
Our neural network takes as input both the galaxy images and the inverse DM mass. 
This method enables scalable, image-based predictions for the theoretical DM SHMFs of individual galaxies, facilitating direct comparisons with observational measurements.
\end{abstract}

\vspace{-10mm}

\section{Introduction}
One of the most striking open questions in modern astrophysics is the nature of dark matter (DM), which constitutes approximately 80\% of the universe’s matter content \citep[e.g.,][]{WMAP_2013, Planck_2020}. While its presence is inferred from gravitational phenomena across a wide range of cosmic scales, from galaxy clusters to large-scale structure, DM has yet to be detected through any non-gravitational interactions, and its fundamental properties remain unknown.

Different models predict distinct clustering behaviors for DM, especially on small, sub-galactic scales. On these scales, the distribution of dark matter — quantified by the subhalo mass function (SHMF) — is highly sensitive to its particle nature, making it a powerful discriminator between DM models \citep[e.g.,][]{Ferreira_2021_DM}. In warm DM (WDM) scenarios, for example, smaller particle masses correspond to higher thermal velocities, which suppress the formation of low-mass halos below the free-streaming scale \citep[e.g.,][]{Colin_2000_WDM, Gilman_2020_HMF, Loudas_2022_WDM}.

Strong gravitational lensing provides a unique way to probe the distribution of matter on these small scales. Unlike methods that rely on luminous tracers, lensing is sensitive to all matter — luminous or dark — making it a powerful observational tool to constrain the SHMF. Traditional analyses infer the presence of individual subhalos by evaluating whether introducing localized perturbers to a smooth lens model leads to a statistically significant improvement in the fit to the observed lensed images \citep[e.g.,][]{Vegetti_2010, Yashar_2016}. More recently, simulation-based studies have demonstrated that machine learning approaches could enable population-level inference of the SHMF by combining data across ensembles of lensing systems \citep[e.g.,][]{brehmer2019mining, Brehmer_Sidd_2019_NRE, Coogan_2022, Zhang_2022, Wagner_C2023, Wagner_C2024, Zhang_2024, OOD_paper_24}.

However, connecting these observational constraints to theoretical predictions remains challenging. Even within a fixed cosmology, the SHMF depends on properties of individual galaxies, including, for example, total mass, morphology, merger history, and local environment \citep[e.g.,][]{Yashar_2016_powerspec}, leading to significant system-to-system variation.

In this work, we introduce a machine learning framework to predict theoretical SHMFs directly from galaxy images, conditioned on an assumed WDM mass. Its goal is to predict, given a WDM particle mass, a plausible theoretical range of SHMFs for specific individual galaxies based on their observable properties. These predictions can then be tested against observational constraints, such as those coming from galaxy-galaxy strong gravitational lensing, to place limits on the WDM mass. To make these predictions, we use the DREAMS simulation suite~\citep{{Jonah_DREAMS_2025}}, as well as \texttt{Synthesizer}~\citep{Vijayan_2020_synthesizer} to create realistic galaxy images. Our method accounts for inter-galaxy variability and enables scalable, image-based inference of theoretical predictions. This approach provides a new pathway to compare dark matter models with forthcoming lensing observations, enabling more precise, per-galaxy tests of DM’s small-scale gravitational effects.

The paper is structured as follows: In Section~\ref{sec:DREAMS}, we describe the hydrodynamical simulation suite used in this work.
Section~\ref{sec:Methods} begins with an overview of how we created the galaxy images from the hydrodynamical simulations, then provides the assumed SHMF profile and the correlation with different DM models, as well as the neural network architecture used to make predictions. Section~\ref{sec:Results} presents our results, and we conclude in Section~\ref{sec:Discussion}.

\section{DREAMS Simulations}
\label{sec:DREAMS}
The DREAMS simulation suite contains, among other products, a set of high-resolution cosmological zoom-in hydrodynamic simulations designed to explore galaxy formation under varying DM and baryonic physics. Each zoom-in simulation is run using the \texttt{AREPO} code \cite{Springel_2010_AREPO, Springel_2019_AREPO, Weinberger_2020_AREPO}, enabling accurate modeling of complex baryonic processes such as gas cooling, star formation, and feedback.
The initial conditions for each zoom-in are constructed by selecting a random, isolated Milky Way–mass halo from a low-resolution volume, then iteratively refining its Lagrangian region using intermediate- and high-resolution particle resampling to define the zoom-in domain \citep[see][]{Jonah_DREAMS_2025}. 

These zoom-in simulations are performed across a range of different WDM models in the range $M_{\rm DM} \in [1.8, 30.3]\, \mathrm{keV}$, sampled uniformly in the inverse $M^{-1}_{\rm DM}$. Further, the supernova (SN) wind, SN energy, and active galactic nuclei (AGN) parameters vary between each zoom-in simulation \cite{Jonah_DREAMS_2025}. The simulations are modeled using the IllustrisTNG baryonic physics prescriptions.

By varying both the DM physics and feedback parameters, the DREAMS simulations allow us to marginalize over baryonic uncertainties and learn a mapping from observable galaxy properties and WDM masses to the underlying SHMF. This marginalization ensures that our model generalizes across different astrophysical scenarios.

The DM sub-halos in the WDM zoom-in simulations are identified using the \texttt{Rockstar} halo finder \cite{Behroozi_2013_Rockstar}. At the time of submitting this work, \texttt{Rockstar} halo catalogs were made available for 815 of the 1024 DREAMS zoom-in simulations. These constitute the labeled dataset used for supervised training of the SHMF inference model.

\section{Methods}
\label{sec:Methods}

\begin{figure}
    \centering
    \includegraphics[width=1.\linewidth]{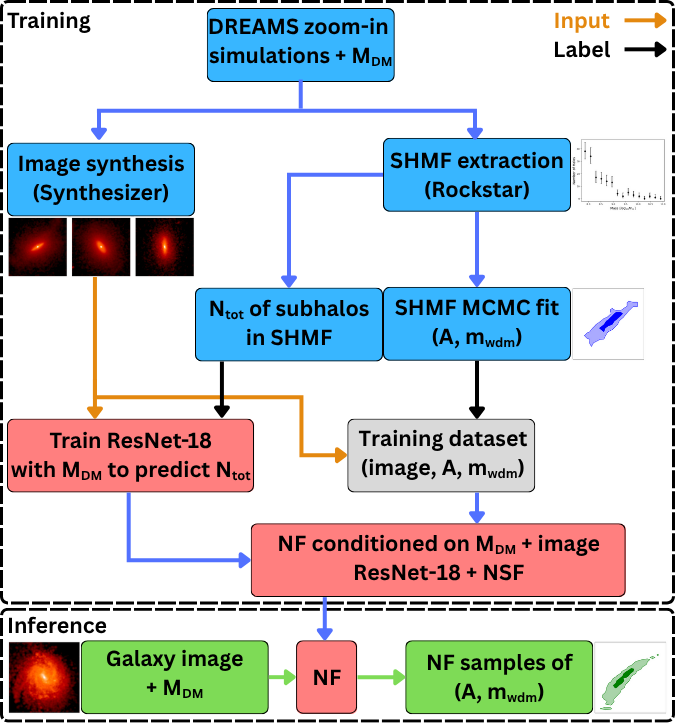}
    \vspace{-8.5mm}
    \caption{\textbf{A flowchart showing the training procedure, label generation, and inference.}}
    \label{fig:flow_architecture}
    \vspace{-5.5mm}
\end{figure}

\subsection{Image Generation with Synthesizer}
\label{sec:Synthesizer}
Creating realistic galaxy images from hydrodynamic simulations typically requires computationally expensive radiative transfer simulations. As an efficient alternative, we use \texttt{Synthesizer} \cite{Wilkins_2020_synthesizer, Vijayan_2020_synthesizer} to generate realistic observational mock images from simulated galaxies.

For each of the 815 zoom-in galaxies with available \texttt{Rockstar}-catalog, we generate 12 different projections by varying the line-of-sight orientation of the particles. \texttt{Synthesizer} produces galaxy images by generating spatially resolved spectral energy distributions (SEDs) and applying instrument-specific wavelength filters from the Spanish Virtual Observatory (SVO) filter service\footnote{\url{https://svo2.cab.inta-csic.es/theory/fps/}} \cite{Rodrigo_2012_SVO, Rodrigo_2020_SVO, Rodrigo_2024_SVO}.

A detailed description of the image creation process can be found in Appendix~\ref{app:synthesizer}.
Examples of the generated galaxy images are shown in Figure~\ref{fig:NF_visual_comp}, alongside their corresponding galaxy-specific SHMFs.

\subsection{Dark Matter Subhalo Mass Function}
Since we aim to learn a mapping between galaxy images and their corresponding SHMF for a given WDM mass, we next move on to modeling the functional form of galaxies' SHMFs. We assume this functional form of the DM SHMFs to be a power-law with a slope of $-0.9$, as an approximation to theoretical predictions of cold DM (CDM) \citep[e.g.,][]{Kuhlen_2007_ViaLactea, Diemand_2007_ViaLactea, Springel_2008, Hiroshima_2018}.
To model the subhalo mass function of WDM, we add to the power-law form of CDM a low-mass cutoff \citep[e.g.,][]{Gilman_2020_HMF}:
\begin{equation}
\begin{split}
    \frac{dN}{dm}|_{\rm wdm} &= \frac{dN}{dm}|_{\rm CDM}  \cdot \left( 1+ \left( \frac{m_{\rm wdm}}{m} \right) \right)^{-1.3} \\ &= A \cdot m^{-0.9} \cdot \left( 1+ \left( \frac{m_{\rm wdm}}{m} \right) \right)^{-1.3}
    \label{eq:pwerlaw_cutoff}
\end{split}
\end{equation}
where $m_{\rm wdm}$ is the cutoff mass, a characteristic of WDM scenarios, and the parameter $A$ is the normalization. 

We fit the parameterized WDM SHMF form to the data of the simulated galaxies, with the subhalos of the individual galaxies identified by \texttt{Rockstar}. 
We bin the identified subhalos in 15 mass bins, linearly spaced in logarithmic 10 base from $\log_{10}(M/M_\odot) = 7.75$ to $11$. 
The uncertainties in the observed subhalo counts $\hat{n}_i$ are dominated by Poisson noise, see Appendix~\ref{app:poisson}. 
To fit the parameters of~\ref{eq:pwerlaw_cutoff}, we use the likelihood defined in eqn~(\ref{eq:likelihood}) in Appendix~\ref{app:poisson}:
\begin{equation}
    \ln\mathcal{L}(\hat{n}\mid\theta) = -\frac{1}{2}\sum_i
     \frac{\bigl(\hat{n}_i - n_i(\theta)\bigr)^2}
          {V_i \;-\; V'_i\,\bigl(\hat{n}_i - n_i(\theta)\bigr)}\, ,
\end{equation}
where the index $i$ runs over the mass bins, $n_i(\theta)$ is the model prediction of eqn~(\ref{eq:pwerlaw_cutoff}), given the parameters $\theta = (A, m_{\mathrm{wdm}})$. 
We use the \texttt{emcee} package \cite{emcee_2013} to perform Markov Chain Monte Carlo (MCMC) sampling over this likelihood. This yields posterior samples for the amplitude $A$ and the WDM cutoff mass $m_{\rm wdm}$.
We then use the samples $(A, m_{\rm wdm})$ obtained in this way to train a normalizing flow (NF), which allows us to account for correlations between $A$ and $m_{\rm wdm}$. Our goal is to then use this NF to predict the expected mass function at a given WDM mass of specific individual galaxies, and compare these predictions to observational constraints from other DM probes, such as strong gravitational lensing.

For each zoom-in galaxy, we use 100 posterior samples from the fitted distribution of $(A, m_{\rm wdm})$ as training labels for the NF. This ensures that the NF learns the continuous distribution of plausible parameters for $A$ and $m_{\rm wdm}$, rather than a single point estimate.
As the flowchart in Figure~\ref{fig:flow_architecture} illustrates, the MCMC samples are only used during training to help the NF learn the continuous distributions of $A$ and $m_{\rm wdm}$ conditioned on the galaxy morphologies and $M_{\rm DM}$.

\subsection{Network Architecture and Training}
Our architecture consists of two main components: a convolutional neural network (CNN) to process the image of the galaxy and a conditional normalizing flow (NF). 

Within a given cosmological model, the predicted SHMF depends on the properties of each individual galaxy.  Given a WDM mass, our architecture infers the relation between the observable properties of the galaxies and their SHMF.
Figure~\ref{fig:flow_architecture} shows a flowchart of the training and inference scheme we used.
We first pre-train a ResNet-18 with a two-layer multi-layer perceptron (MLP) to predict the number of subhalos in the galaxy images, with the inverse WDM mass $M^{-1}_{\rm DM}$ concatenated to the first of the two MLP layers, using a mean squared error (MSE) loss. Our pretraining ensures a meaningful embedding space for the images, before using the output of the pretrained CNN as a condition for the NF. 

The embedded image and the inverse of the WDM mass $M^{-1}_{\rm DM}$ are used as conditions for a neural spline flow (NSF) to predict the parameters $A$ and $m_{\text{WDM}}$. 
We use the \texttt{zuko} library to implement the NSF \cite{Rozet_2022_zuko}. We use 730 of the zoom-in galaxies as training data and keep 85 as validation data, for both - the pretraining of the CNN and the training of the NSF - the same sets. 

To train the full model, all the weights of the ResNet are allowed to update, enabling end-to-end optimization of the feature extractor and density estimator. This allows us to efficiently model the posterior over the SHMF fit parameters conditioned on both the image and the WDM mass.

\section{Results}
\label{sec:Results}
\begin{figure*}[th]
    \centering
    \vspace{-3mm}
    \includegraphics[width=0.33\textwidth]{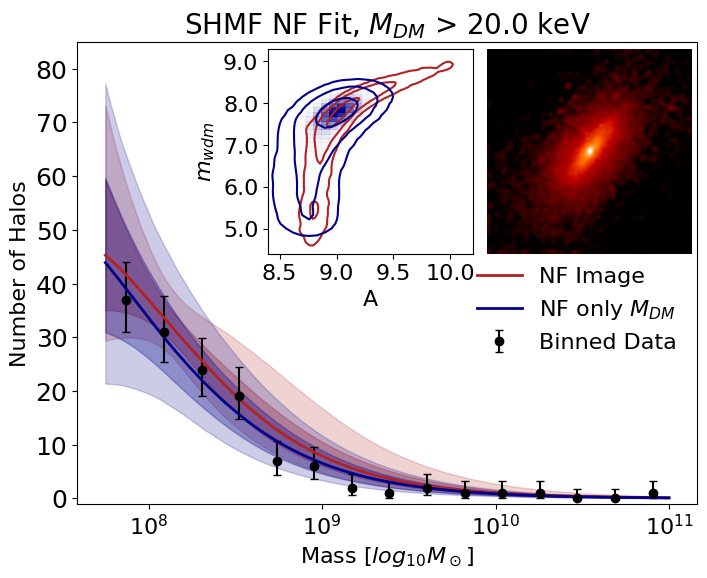}
    \includegraphics[width=0.33\textwidth]{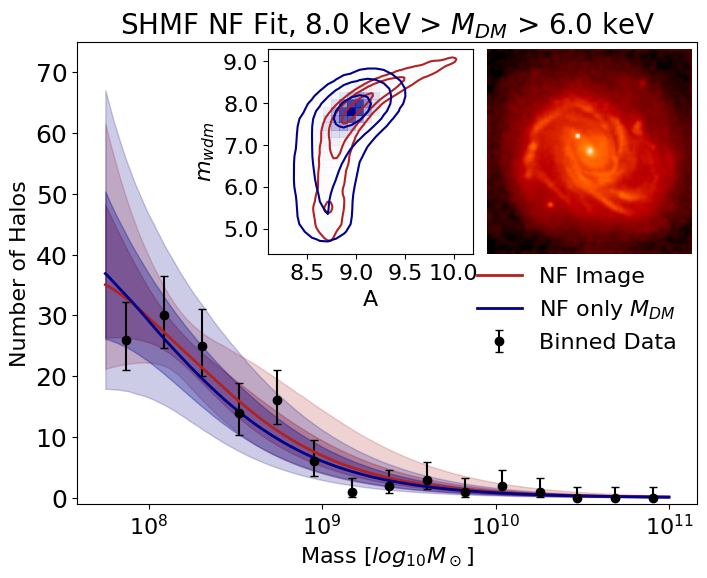}
    \includegraphics[width=0.33\textwidth]{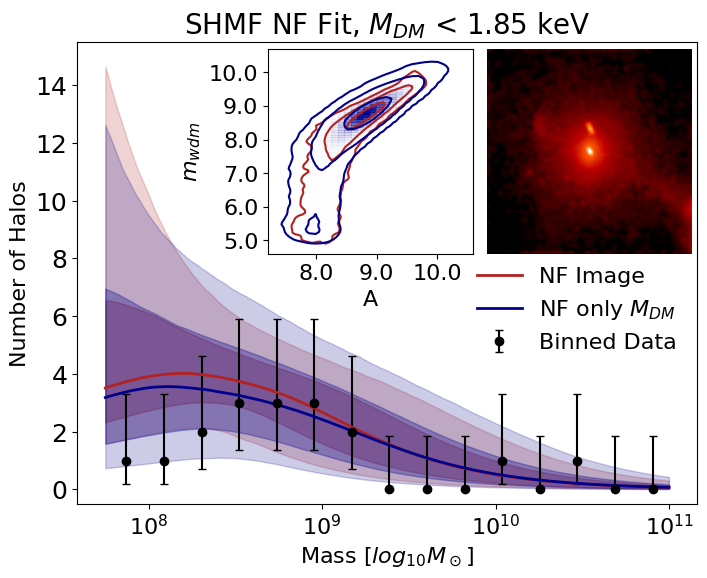}
    \includegraphics[width=0.33\textwidth]{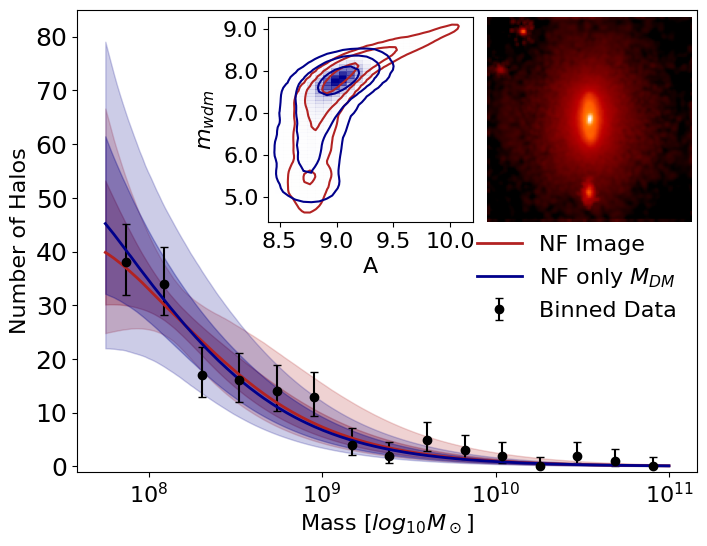}
    \includegraphics[width=0.33\textwidth]{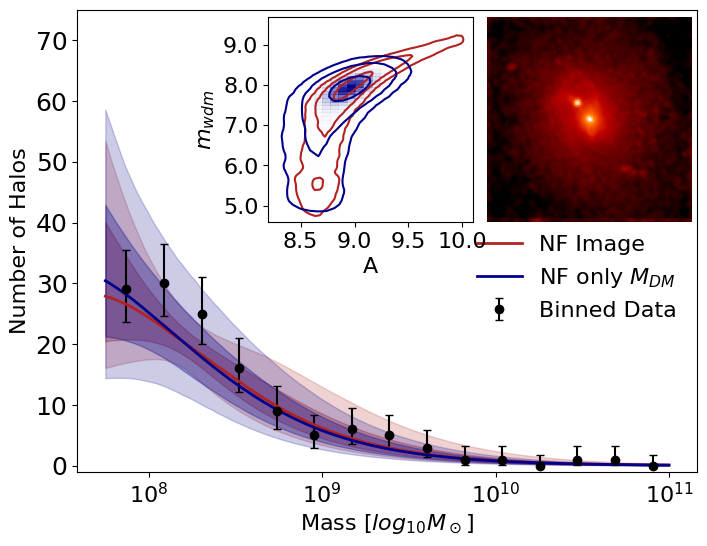}
    \includegraphics[width=0.33\textwidth]{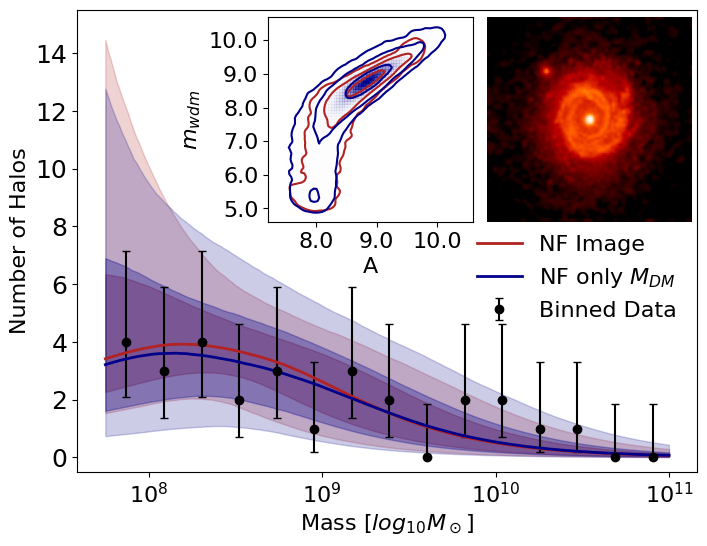}
    \vspace{-8.5mm}
    \caption{\textbf{The galaxy-specific halo mass functions under different cosmologies.} The figure shows the galaxy-specific halo mass functions for different WDM masses, with the corresponding galaxy image in the plot. The displayed galaxies come from three different WDM mass regions. The black dots are the counts of subhalos from the DREAMS simulation with Poisson uncertainties.
    The blue contours show the samples of the NF conditioned only on $M_{\rm DM}$, and the red contours the samples of the NF conditioned on both $M_{\rm DM}$ and the galaxy image. The NF conditioned only on $M_{\rm DM}$ shows a broader posterior range and bigger uncertainties on the fit.}
    \label{fig:NF_visual_comp}
    \vspace{-5mm}
\end{figure*}

Figure~\ref{fig:NF_visual_comp} shows galaxy-specific SHMFs with the corresponding galaxy image.
We compare the NF conditioned on the galaxy images and $M_{\rm DM}$, and an NF conditioned only on $M_{\rm DM}$. This gives an estimate of how including the morphological information of the galaxy can improve the SHMF prediction, when compared to sampling a SHMF that has been marginalized over galaxy morphologies, as usually assumed from theoretical predictions.
Since each galaxy was only simulated for a single WDM temperature in the DREAMS suite, Figure~\ref{fig:NF_visual_comp} performs this comparison for galaxies from the test set at a single WDM mass for each galaxy.
We achieve an improvement in the SHMF predictions by including information about the galaxy morphology.
The displayed galaxies come from three different WDM mass regions. 
The examples on the left come from higher WDM masses, the ones in the middle from medium masses, and the right side from low WDM masses.
The black dots are the subhalo counts from the DREAMS simulations with Poisson uncertainties. 
The red and blue solid lines show the 50th percentile fit of the NF samples conditioned on the image and the samples of the NF only conditioned on $M_{\rm DM}$, respectively. The shaded regions show the $1\sigma$ and $2\sigma$ fits of the SHMF from eqn~\ref{eq:pwerlaw_cutoff} using the sampled parameters.
The displayed contours show the $1\sigma$, $2\sigma$, and $3\sigma$ contours of the parameter samples $(A, m_{\rm wdm})$. The titles indicate the WDM mass range, which is passed along with the image to the neural network. For each range of $M_{\rm DM}$, we show two examples in the same column.
The NF conditioned only on $M_{\rm DM}$ shows broader posteriors and larger uncertainties on the parameters of the SHMF. On the other hand, including the image information allows tighter constraints on both parameters, $A$ and $m_{\rm wdm}$, and highlights their correlations, resulting in tighter constraints on the SHMF. This supports the conclusion that incorporating galaxy images enables more precise predictions of the SHMF fit parameters than using the WDM mass $M_{\rm DM}$ alone.

\begin{table}[t]
    \centering
    \vspace{-3mm}
    \caption{\textbf{PQMass-$\chi^2$ and RSME values}. The table shows the mean $\chi^2$ values obtained with PQMass, as well as the median RSME, for the NF conditioned on images and $M_{\rm DM}$, and an NF conditioned only on $M_{\rm DM}$ for the entire validation set. The comparison is made between the MCMC samples and the different NF samples for $A$ and $m_{\rm wdm}$.}
    \vspace{1mm}
    \begin{tabular}{l|c|c}
            \hline \hline
         & NF with image & NF without image \\ \hline
         PQM $\chi^2$ & $343.7\pm92.5$ & $385.8 \pm 121.7$ \\ 
         Median RMSE & $0.26\pm 0.20$ & $0.55\pm0.17$ \\
         \hline \hline
    \end{tabular}
    \vspace{-6mm}
    \label{tab:PQM}
\end{table}

We further test the NF conditioned on the galaxy images and $M_{\rm DM}$ against the NF only conditioned on $M_{\rm DM}$ qualitatively, to show that the inclusion of the galaxy images leads to a better performance of the NF. 
In Table~\ref{tab:PQM} we report the mean and standard deviation of the PQMass-$\chi^2$ values \citep{PQMass} for the entire validation set for the NF conditioned on images and $M_{\rm DM}$, and the NF only conditioned on $M_{\rm DM}$. We also report the median root mean squared error (RMSE) and its standard deviation of the entire validation set for both network cases.
For that, we use for each condition the MCMC samples of the SHMF fit as target distribution and compare those against the NF samples of $A$ and $m_{\rm wdm}$. The closer the distributions are, the lower the PQMass-$\chi^2$ value. 
We do not expect the NF samples to match the MCMC fits perfectly, because each individual galaxy samples a distribution of possible SHMF for a given WDM model, making the inference under-specified without additional information. Providing additional galaxy images leads to a more informative inference of the possible SHMF, since they provide more constraining information.
The lower RMSE value for the normalizing flow with access to the morphological information shows that the performance of the architecture improves and has tighter constraints on the SHMF.
Additionally, we compare the MCMC fits of the SHMFs and the NF samples of the NF with image conditioning in Figure~\ref{fig:hmf_fits} in Appendix~\ref{app:NF_cond}. We do not compare the NF samples here with the MCMC samples, because the goal is to show that the NF with access to the image data does outperform the pure theoretical NF predictions based on the WDM only without image access.

The current sample size of the simulation is too small to have a separate test set; therefore, we show the performance for the validation set.
The provided examples show that the network is able to learn the galaxy-specific fit parameters within the predicted uncertainties.

\section{Discussion}
\label{sec:Discussion}
Future iterations of this work will focus on increasing the realism of the input images and expanding the training dataset. The current images do not include the effects of dust attenuation, assume only Gaussian noise, and use a simplified Gaussian point spread function (PSF). Additionally, the images are expressed in flux units ($\mathrm{erg,s^{-1},Hz^{-1}}$) rather than in instrument-specific units. We plan to convert them to the native units of the Hubble Space Telescope (HST), i.e., $\mathrm{e^{-},s^{-1}}$, and to incorporate non-Gaussian, instrument-specific noise using SLIC \citep{Ronan_2023_SLIC}, along with more realistic PSF models. These improvements will be accompanied by the inclusion of the full set of \texttt{Rockstar} halo catalogs, including those for the remaining 209 DREAMS zoom-in simulations, to expand the training and validation sets and enable evaluation on an independent test set.

The results demonstrate strong performance and highlight the potential of applying this architecture to more realistic simulations. However, there are important limitations to note regarding the training data. All galaxies used during training and validation are Milky Way-mass halos in isolated systems. This constraint is due to the availability of high-resolution DM zoom-in simulations, which currently exist only for these galaxy types, since running such detailed zoom-ins on cluster galaxies is computationally too expensive. Whilst the masses of the galaxies are for all galaxies in the order of the Milky Way, they are approximately equally distributed between spiral and elliptical galaxies.
In contrast, strong gravitational lenses are more commonly observed to be massive galaxies in dense environments - often the central galaxies of clusters. These systems are not represented in the necessary mass resolutions in current hydrodynamical simulations. For these purposes, the subhalos should be resolved down to masses of $\log_{10}(M/M_\odot) = 7.75$ and below to be able to resolve small subhalos and confidently infer the SHMF.

We expect that future high-resolution hydrodynamical simulations will model the relevant environments in greater detail, allowing us to extend the training set and improve generalization to observed lensing galaxies. Until then, our results should be interpreted within this limitation.

\section*{Acknowledgements}
This work is partially supported by Schmidt Sciences, a philanthropic initiative founded by Eric and Wendy Schmidt as part of the Virtual Institute for Astrophysics (VIA). The work is in part supported by computational resources provided by Calcul Quebec and the Digital Research Alliance of Canada. A.F. acknowledges the support from the Bourse J. Armand Bombardier and UdeM's final year scholarship.
T.N. is supported by the CIERA Postdoctoral Fellowship. Y.H. and L.P. acknowledge support from the Canada Research Chairs Program, the National Sciences and Engineering Council of Canada through grants RGPIN-2020-05073 and 05102.

\bibliography{bibliography}
\bibliographystyle{icml2025}

\newpage
\appendix
\onecolumn

\section{Image Generation with Synthesizer}
\label{app:synthesizer}
\texttt{Synthesizer} produces galaxy images by generating spatially resolved spectral energy distributions (SEDs) and applying instrument-specific wavelength filters from the Spanish Virtual Observatory (SVO) filter service\footnote{\url{https://svo2.cab.inta-csic.es/theory/fps/}} \cite{Rodrigo_2012_SVO, Rodrigo_2020_SVO, Rodrigo_2024_SVO}.
We use the Hubble Space Telescope Wide Field Camera (HST/WFC) F105W filter to simulate realistic near-infrared observations.

The synthetic images are constructed by only taking stellar particles into account. We neglect the contributions of dust and gas particles, which are expected to have a relatively minor impact on the morphological features relevant to our analysis. In future steps, we will include the effects of dust and interstellar gas.
As the first step of the image creation, we use an incident emission model and place the galaxies at redshift $z=0$. 

The particles are convolved with a smoothing filter for a more optical appealing visualization.
The smoothing length of the individual stellar particles is taken for each particle from the simulation data directly, which means that any artifacts of isolated particles in the imaging are originating from the simulation. The smoothing length is the co-moving radius of the sphere centered on the particle enclosing the $32\pm1$ nearest particles of this same type.

To ensure image fidelity, we first generate high-resolution images and convolve them at the high resolution with a Gaussian point spread function (PSF) to approximate observational effects. 
The PSF has a full-width half maximum of 3 pixels. 
The resulting images are then downsampled to a target resolution of 0.78125 kpc/pixel.
We add gaussian noise of $10^{23} \frac{\rm erg}{\rm s \cdot Hz}$. We do not yet include instrument-specific observational noise or Poisson noise.
Examples of the generated galaxy images are shown in Figure~\ref{fig:NF_visual_comp}, alongside the corresponding galaxy-specific SHMFs.

\section{Poisson Noise from Samples}
\label{app:poisson}
To compute the Poisson noise of the counts of each bin of the SHMF, we use the \texttt{Rockstar} catalogs. We follow the prescription in \citet{Tanabashi_2018_Poisson, Chang_2021_poissonoise}. We compute the confidence intervals using the inverse cumulative distribution function of the $\chi^2$ distribution:
\begin{align}
    \mu_{\rm lower} &= \frac{1}{2}\,F^{-1}_{\chi^2}\!\Bigl(\tfrac{\alpha}{2};\,2\hat{n}\Bigr) \\
    \mu_{\rm higher} &= \frac{1}{2}\,F^{-1}_{\chi^2}\!\Bigl(1-\tfrac{\alpha}{2};\,2(\hat{n}+1)\Bigr)
    \label{eq:poisson_uncert}
\end{align}
with $F^{-1}_{\chi^2}$ the inverse of the $\chi^2$ cumulative distribution and $\hat{n}$ the number of counts. The confidence level is given by $100(1-\alpha)$.
From these intervals, we define the asymmetric error bars and variances with:
\begin{align}
    \sigma_{\mathrm{lower},i} &= \hat{n}_i \;-\;\mu_{\mathrm{lower},i}\\
    \sigma_{\mathrm{higher},i} &= \mu_{\mathrm{higher},i}\;-\;\hat{n}_i\\
    V_i &= \sigma_{\mathrm{lower},i}\,\sigma_{\mathrm{higher},i}
    \label{eq:v_i}
    \\
    V'_i &= \sigma_{\mathrm{higher},i} \;-\;\sigma_{\mathrm{lower},i}
    \label{eq:V_i_prime}
\end{align}
We use the variances and uncertainties on the counts per bin to compute the likelihood of our fits to the data.

The likelihood function to fit the SHMF (eqn~\ref{eq:pwerlaw_cutoff}) is defined by:
\begin{equation}
    \ln\mathcal{L}(\hat{n}\mid\theta) = -\frac{1}{2}\sum_i
     \frac{\bigl(\hat{n}_i - n_i(\theta)\bigr)^2}
          {V_i \;-\; V'_i\,\bigl(\hat{n}_i - n_i(\theta)\bigr)}
    \label{eq:likelihood}
\end{equation}
where $n_i(\theta)$ is the model prediction of eqn~(\ref{eq:pwerlaw_cutoff}), given the parameters $\theta = (A, m_{\mathrm{wdm}})$. 
We use the likelihood function to fit the \texttt{Rockstar}-catalog data with an MCMC.

\section{Comparison of Normalizing Flow Samples with MCMC Fits}
\label{app:NF_cond}
\begin{figure*}[ht]
    \centering
    \includegraphics[width=0.33\textwidth]{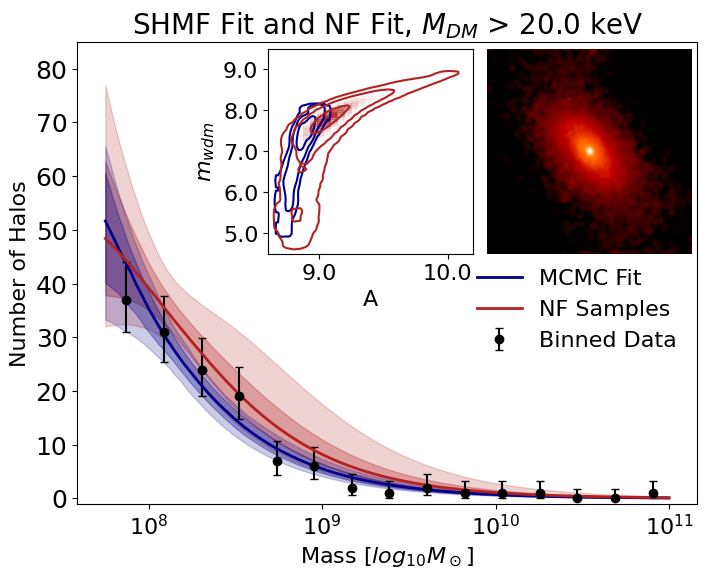}
    \includegraphics[width=0.33\textwidth]{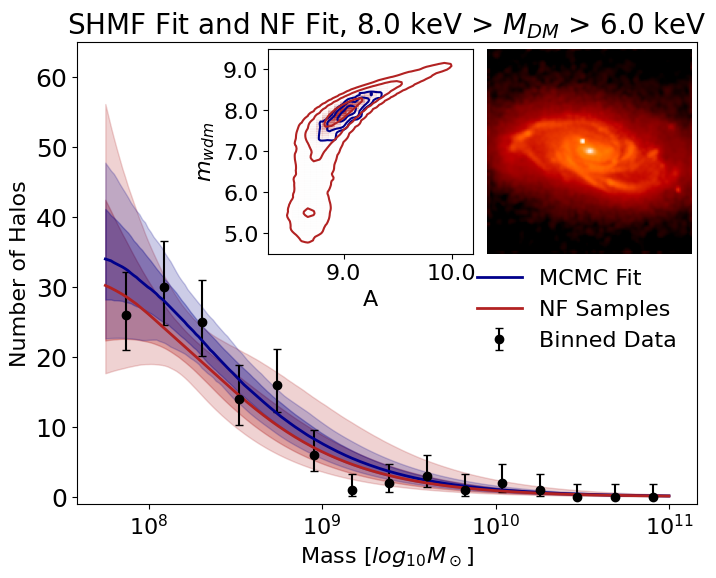}
    \includegraphics[width=0.33\textwidth]{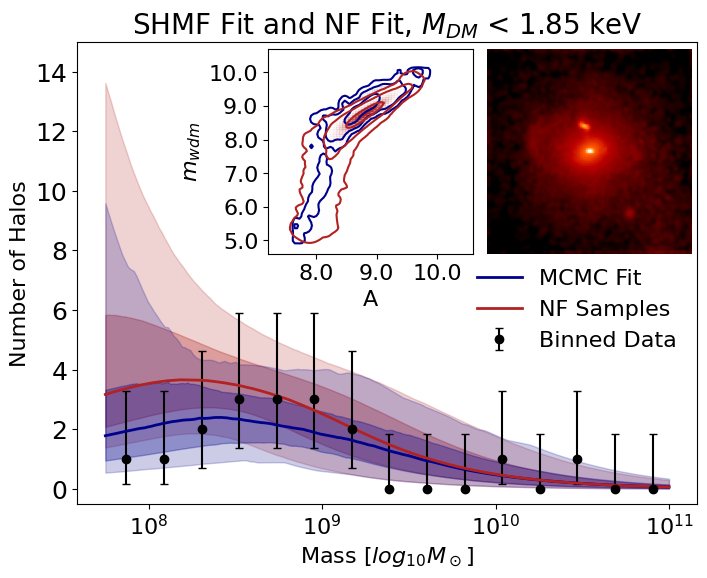}
    \includegraphics[width=0.33\textwidth]{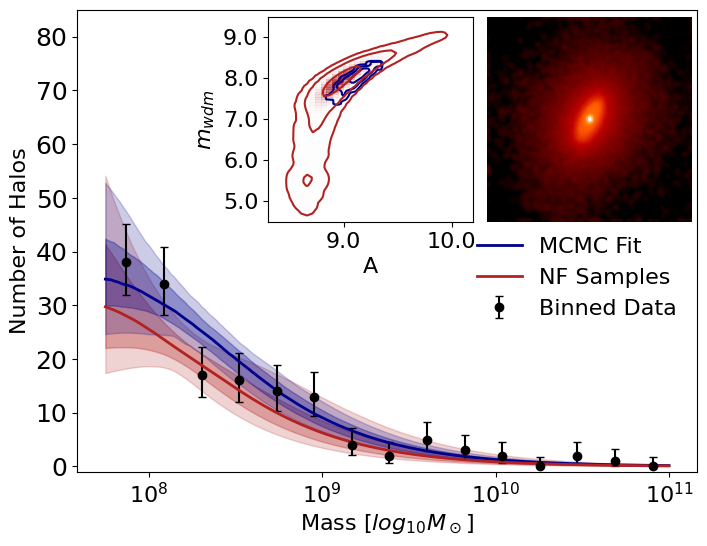}
    \includegraphics[width=0.33\textwidth]{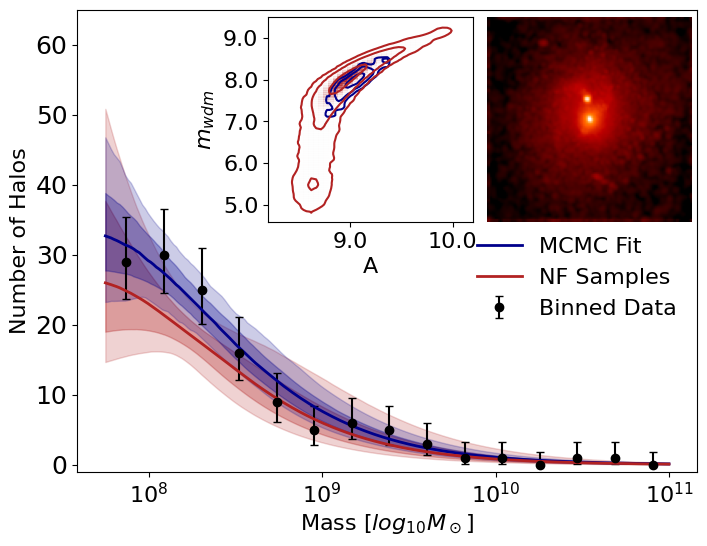}
    \includegraphics[width=0.33\textwidth]{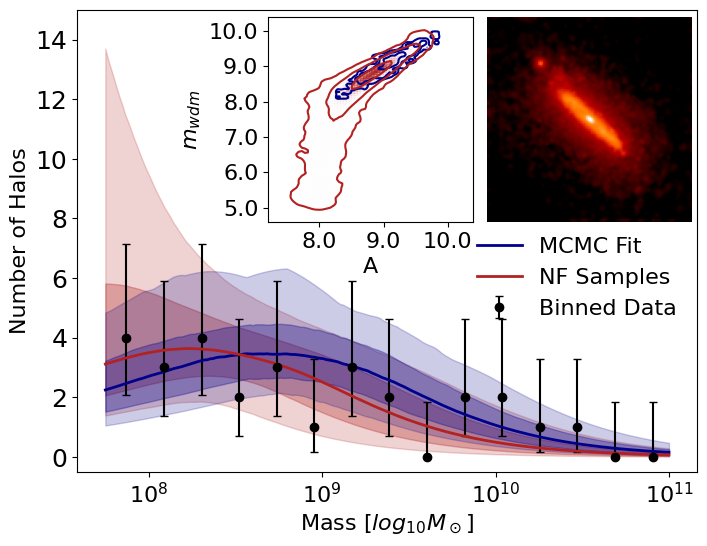}
    \caption{\textbf{Similar to Figure~\ref{fig:NF_visual_comp}.} The shown galaxies are the same as in Figure~\ref{fig:hmf_fits}, but the orientation of the galaxy images is different. The blue contours show the samples of the MCMC fit, and the red contours the samples of the NF conditioned on $M_{\rm DM}$ and the galaxy image. Overall, the sampled parameters of the NF represent the fits obtained directly through the likelihood and cover the full range of MCMC sampled parameters.
    }
    \label{fig:hmf_fits}
\end{figure*}

In Figure~\ref{fig:hmf_fits}, we show galaxy-specific SHMFs obtained by NF samples and MCMC fits along with the corresponding galaxy image. The figure is structured as Figure~\ref{fig:NF_visual_comp}. We use the same galaxies as before, but under different projections.
We do not expect to perfectly recreate the MCMC fits. Our method serves as a tool to get the theoretical predictions on the SHMF given a WDM theory. This does not serve as constraints on the SHMF, but is rather to be used to compare theoretical predictions with observational constraints that can be achieved with strong lensing results. This is not a standalone method to learn about dark matter in observational surveys, but it needs a comparison counterpart that constrains the SHMF. 
The blue and red solid contours show the samples of the MCMC fit and of the NF samples, respectively. The dark and light shaded areas are the $1\sigma$ and $2\sigma$ regions of the fits. The displayed contours show the $1\sigma$, $2\sigma$, and $3\sigma$ contours of the MCMC fits and NF samples of $(A, m_{\rm wdm})$, defining the SHMF. 

The sampled parameters of the NF represent the fits obtained directly through the likelihood and cover the full range of the parameters sampled with MCMC.


\end{document}